\documentclass[aps,prl,floatfix,showpacs, twocolumn,preprintnumbers,amsmath,amssymb,superscriptaddress]{revtex4}

\usepackage[pdftex]{graphicx}
\usepackage{dcolumn}   
\usepackage{bm}        
\usepackage{rotating}  

\begin{document}

\title{Magnetic control of spin-orbit fields: a first-principles study of Fe/GaAs junctions}
\author{Martin Gmitra}
\affiliation{Institute for Theoretical Physics, University of Regensburg, 93040 Regensburg, Germany}
\author{Alex Matos-Abiague}
\affiliation{Institute for Theoretical Physics, University of Regensburg, 93040 Regensburg, Germany}
\author{Claudia Draxl}
\affiliation{Physics Department, Humboldt-Universit\"{a}t zu  Berlin,  12489 Berlin, Germany}
\author{Jaroslav Fabian}
\affiliation{Institute for Theoretical Physics, University of Regensburg, 93040 Regensburg, Germany}


\begin{abstract}
The microscopic structure of spin-orbit fields for the technologically important Fe/GaAs 
interface is uncovered from first principles. A symmetry based method allows to obtain
the spin-orbit fields---both their magnitude and orientation---for a generic Bloch state, 
from the electronic band structure for any in-plane magnetization orientation.  It  is
demonstrated that the spin-orbit fields depend not only on the electric field across the
interface, but also surprisingly strongly on the Fe magnetization orientation, opening
prospects for their magnetic control. These results give important clues 
in searching for spin-orbit transport and optical phenomena in ferromagnetic/nonmagnetic systems.
\end{abstract}
\maketitle

In solid-state systems lacking space inversion symmetry spin-orbit
coupling (SOC) acts on the electronic structure as a spin-orbit field (SOF),
which is an effective magnetic field whose direction and magnitude
depend on the electron
momentum \cite{Zutic2004:RMP, Fabian2007:APS}.
The most prominent examples are the Dresselhaus spin-orbit
field \cite{Dresselhaus1955:PR} describing the effects of bulk
inversion asymmetry (BIA) in zinc-blende semiconductors, and
the Bychkov-Rashba spin-orbit field \cite{Bychkov1984:JETPL},
describing the effects of structure inversion asymmetry (SIA) in
asymmetric quantum wells. Apart from semiconductor structures, where
Bychkov-Rashba coupling has been extensively studied
\cite{Winkler2003:book,Fabian2007:APS,Cartoixa2006:PRB}
it has been investigated in many other systems, for example on
metallic surfaces
\cite{LaShell1996:PRL, Henk2003:PRB, Koroteev2004:PRL,
Fluegel2006:PRL,Ast2007:PRL,Meier2008:PRB}, graphene
on a Ni substrate \cite{Dedkov2008:PRL}, or in Au and Ag
monolayers on W(110) substrates \cite{Shikin2008:PRL}.
A striking manifestation of spin-orbit coupling in condensed matter
is the spin-momentum locking in topological
insulators \cite{Hasan2010:RMP}.

Spin-orbit coupling can be controlled by an electric field \cite{Nitta1997:PRL}. This fact
has for long been used to motivate spintronics applications as epitomized by the
Datta-Das transistor \cite{Datta1990:APL} in which the gate controls the spin-orbit induced spin
precession of the itinerant electrons in a transistor channel. But spin-orbit coupling is
also important for anisotropic magnetotransport. Tunneling
anisotropic magnetoresistance (TAMR), for example, can be used
to control electrical transport by rotating the magnetization orientation
of a single ferromagnetic layer. It has been observed and studied in a variety of systems,
{GaMnAs/Al} \cite{Gould2004:PRL}, Fe/GaAs, \cite{Moser2007:PRL, Fabian2007:APS},
CoFe/GaAs  \cite{Uemura2010:APL} (inserting an MgO barrier suppresses TAMR here \cite{Akiho2011:APL},
a clear evidence for interface induced symmetry of the effect), 
Co/Pt \cite{Park2008:PRL},
Si/ferromagnet junctions \cite{Sharma2012:PRB},
resonant tunnel devices \cite{Tran2009:APL},
or on an atomic scale in STM experiments \cite{Bergmann2012:PRB}.
Interfacial spin-orbit coupling has been proposed to control thermoelectric
anisotropies in helimagnetic tunnel junctions \cite{Jia2011:APL} and
produce spin-transfer torque in ferromagnet-topological
insulator junctions \cite{Mahfouzi2012:PRL}.

In earlier studies of spin-orbit coupling  on surfaces \cite{LaShell1996:PRL,
Henk2003:PRB,Koroteev2004:PRL,Krupin2005:PRB,Ast2007:PRL,
Meier2008:PRB} and interfaces \cite{Dedkov2008:PRL,Shikin2008:PRL}
the spin-orbit Hamiltonian was extracted by fitting the energy bands
close to the $\Gamma$ point assuming a Bychkov-Rashba-type coupling.
This standard procedure requires {\em a priori} knowledge of the specific
functional form of the spin-orbit field and applies only to very
small $\mathbf{k}$-vectors for which small-momentum expansions are meaningful.
Here we introduce a novel method to obtain spin-orbit fields (not just the functional
parameters) for a generic
 $\mathbf{k}$-point directly from {\em ab-inito} data. On the example of an
Fe/GaAs junction, important for room temperature spin injection \cite{Hanbicki2002:APL,Crooker2005:S,Kotissek2007:NP,%
Mavropoulos2002:PRB,Perlov2002:PSSB} and TAMR \cite{Moser2007:PRL, Sykora2012:JPCM},
we derive a formula for the magnitude and direction of the momentum dependent
spin-orbit fields directly form the electronic band structure.
The results show highly anisotropic (with respect to the momentum
orientation) patterns, which take on
different forms, from the ones known in semiconductor physics for
small momenta  to more exotic ones for Bloch states further away
from the $\Gamma$ point.

One fascinating outcome is a qualitative dependence of the spin-orbit fields patterns
on the band (energy), consistent with the bias-induced inversion of the TAMR
observed in experiments \cite{Moser2007:PRL,Matos-Abiague2009:PRB}.
Even more important, in addition to their sensitivity on an electric field,
the spin-orbit fields can depend unusually
strongly on the magnetization direction, to the point that the anisotropy axes can be
flipped by rotating the magnetization. We emphasize that those
effects are caused by the symmetry of the interface, not of
the bulk structures,  making them particularly important for
lateral transport in ultrathin hybrid ferromagnet-nonmagnet junctions.

We consider thin Fe/GaAs slabs. The small lattice mismatch between
Fe ($2.87\,{\rm \AA}$) and GaAs
($5.65\,{\rm \AA}$) allows for a smooth epitaxial growth of Fe on
a GaAs (001) surface. Early investigations of the stability of
$1\times 1$ Fe/GaAs interfaces within density functional theory
\cite{Erwin2002:PRB} showed that when more than two atomic layers
of Fe are deposited on a GaAs (001) surface, the flat or partially
intermixed interfaces are more stable than the fully intermixed one,
the As-terminated flat interface being more stable that the partially
intermixed one.
On the other hand, a recent Z-contrast scanning transmission electron microscopy reported
a single plane of alternating Fe and As atoms at an Fe/AlGaAs interface
\cite{Zega2006:PRL, Fleet2011:JAP}. Since the choice of the interface
is not important to the message of our
paper, we choose an As-terminated flat interface.

The electronic structure of an ideal Fe/GaAs slab, containing 9 (001) atomic
layers of GaAs with the diagonal lattice spacing $d=a/\sqrt{2}=3.997\,{\rm \AA}$
and three atomic planes of bcc Fe, has been calculated
using the full potential linearized augmented plane wave technique
implemented in the FLEUR code \cite{fleur} and a generalized gradient
approximation for the exchange-correlation functional \cite{Perdew1996:PRL}.
The SOC has been treated
within the second variational method.

\begin{figure}[!h]
\centering
\includegraphics[width=0.9\columnwidth,angle=0]{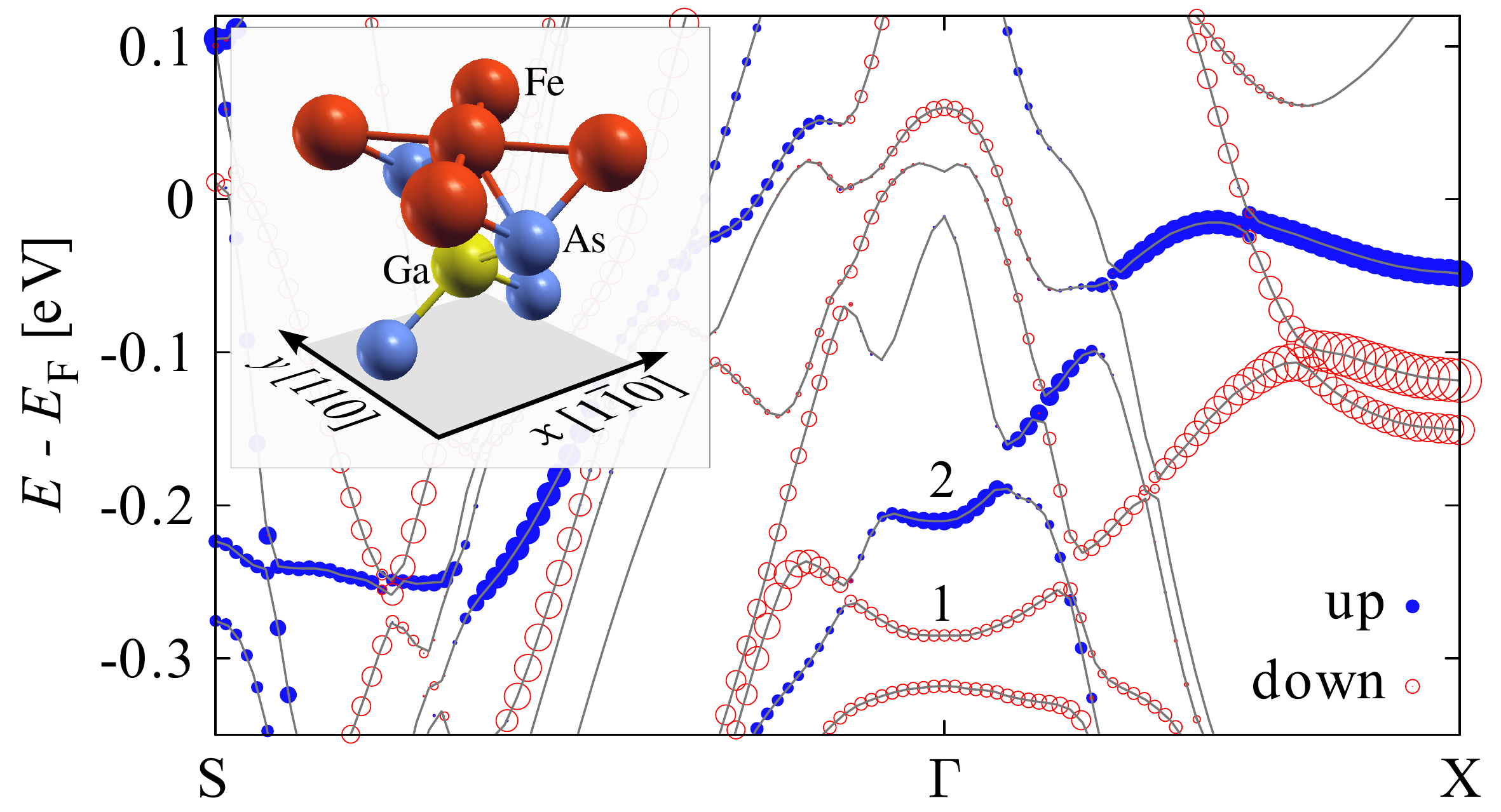}
\caption{{Calculated band structure for the Fe/GaAs slab
and magnetization along $[1\bar{1}0]$.}
The states with spin-up (spin-down) character at the Fe/GaAs interface are
emphasized by blue filled (red open) circles whose radii are
proportional to the corresponding charge density at the interface
atoms. The inset shows the As-terminated flat $1\times 1$ interface
model assumed in the study.
}
\label{Fig:bands}
\end{figure}

The band structure of the Fe/GaAs slab along the high symmetry
lines connecting the ${\rm S - \Gamma - X}$ points in the Brillouin zone
(BZ) is shown in Fig.~\ref{Fig:bands} for a magnetization orientation
along the $\rm [1\bar{1}0]$ direction. The spin character of bands 1 and 2 in Fig.~\ref{Fig:bands} is basically determined by the interface atoms.
The interface unit cell contains interfacial As, the neighboring Ga, and two Fe atoms. The spin-up
character of band $n=2$ is dominated by the interfacial As atom, its neighboring Ga atom
and  Fe atom above  Ga, while the spin-down character of band $n=1$ comes mostly
from the two Fe atoms.

The non-centrosymmetric GaAs layer is of
$D_{2d}$ symmetry, exhibiting the BIA spin-orbit coupling.
The interface lowers the symmetry to $C_{2v}$ with the twofold
rotation axis $C_2$ along the growth direction [001] \cite{Fabian2007:APS}.
The $C_{2v}$ symmetry accounts for both the BIA and SIA; the $C_{2v}$
spin-orbit field lies in the plane of the slab, perpendicular to the
growth direction. Since $C_{2v}$ symmetry has only one-dimensional orbital
irreducible representations, away from accidental level (anti)crossings
the spin-orbit fields (even at high symmetry points) can be described
by spin $1/2$ Pauli matrices.

 The most general SOC Hamiltonian consistent with $C_{2v}$ symmetry
can be written for the in-plane momenta around the $\Gamma$ point as
\begin{equation}\label{Eq:Ham}
{\cal H}_{\rm so}= \mu_{n}(k_x,k_y,\theta) k_x \sigma_y +
\eta_{n}(k_x,k_y,\theta) k_y \sigma_x \,,
\end{equation}
where $k_x$ and $k_y$ are the components of the in-plane wave vector $\mathbf{k}$,
$\sigma_x$ and $\sigma_y$ are the Pauli matrices, and $x$ and $y$ correspond
to the diagonal ${\rm [1{\bar 1}0]}$ and ${\rm [110]}$ crystallographic
directions in GaAs, respectively; $\theta$ refers to the magnetization direction
with respect to the $[1\bar{1}0]$ crystallographic direction of GaAs and $n$ labels the band of interest.
The functional parameters $\mu_{n}$ and $\eta_{n}$,
\begin{flalign}\label{a-so-expa}
& \mu_{n}(k_x,k_y,\theta) = \mu^{(0)}_{n}(\theta) + \mu^{(1)}_{n}(\theta) k_x^2 + \mu^{(2)}_{n}(\theta) k_y^2 + \ldots \,,
\nonumber\\
& \eta_{n}(k_x,k_y,\theta) = \eta^{(0)}_{n}(\theta) + \eta^{(1)}_{n}(\theta) k_x^2 + \eta^{(2)}_{n}(\theta) k_y^2 + \ldots
\end{flalign}
are even in the momenta and, what is crucial and new here, depend in general on the magnetization direction.

The values of the expansion parameters $\mu^{(i)}_{n}$, $\eta^{(i)}_{n}$
($i=0,1,2,\ldots$) determine the specific form of the SOF. For example,
if $\mu^{(0)}_{n}=\alpha_{n}$ and $\eta^{(0)}_{n}=-\alpha_{n}$   ($\mu^{(0)}_{n}=\eta^{(0)}_{n}=\beta_{n}$),
${\cal H}_{\rm so}$ reduces in the limit of small $k=|\mathbf{k}|$ to the
well known Bychkov-Rashba \cite{Bychkov1984:JETPL} (linearized Dresselhaus
\cite{Dresselhaus1955:PR}) SOC with $\alpha_n$ ($\beta_n$) denoting the Bychkov-Rashba (Dresselhaus) SOC parameter of the $n$th band.
By introducing the SOF field
\begin{equation}\label{a-sof-def}
\bm{w}_{n}(k_x,k_y,\theta) = \left( \begin{array}{c}
\eta_{n}(k_x,k_y,\theta) k_y \\
\mu_{n}(k_x,k_y,\theta) k_x \\
0
\end{array} \right) ,
\end{equation}
Eq. (\ref{Eq:Ham}) can be rewritten as ${\cal H}_{\rm so} =
\bm{w}_{n}(\mathbf{k})\cdot\bm{\sigma}$,
where $\bm{\sigma}$ is the vector of the Pauli matrices.

We first analyze the spin-orbit fields for in-plane magnetization directions.
Since the exchange field dominates over spin-orbit coupling and the magnetization lies
in the plane of the layers, the SOC contribution to the energy can be treated
within first order perturbation theory. From the symmetry properties
we find (see the Supplementary Material for the details) the following relations,
\begin{equation}\label{a-wx}
   w_{nx}(\mathbf{k},\theta)=\sigma\left[\frac{\Delta E_{n}^{\rm so}(\mathbf{k},\theta)+\Gamma_{n}^{\rm so}(\mathbf{k},\theta)}{2\cos\theta}\right]
\end{equation}
and
\begin{equation}\label{a-wy}
    w_{ny}(\mathbf{k},\theta)=\sigma\left[\frac{\Delta E_{n}^{\rm so}(\mathbf{k},\theta)-\Gamma_{n}^{\rm so}(\mathbf{k},\theta)}{2\sin\theta}\right]
\end{equation}
where
\begin{equation}\label{a-de-so-fp}
    \Delta E_{n}^{\rm so}(\mathbf{k},\theta)=\frac{E_{n}(\mathbf{k},\theta)-E_{n}(-\mathbf{k},\theta)}{2},
\end{equation}
\begin{equation}\label{a-de-so-g}
    \Gamma_{n}^{\rm so}(\mathbf{k},\theta)=\frac{E_{n}(-k_{x},k_{y},\theta)-E_{n}(k_{x},-k_{y},\theta)}{2},
\end{equation}
and $\sigma$ refers to the spin character of the $n$th band. The above relations allow us to extract the components of the SOF directly from the \emph{ab-initio} energy bands. In the particular cases of $\theta \approx  \pi/2$ and $\theta \approx  0$ the numerators and denominators in Eqs.~(\ref{a-wx}) and (\ref{a-wy}), respectively, vanish.
In such cases the SOF is obtained by L'H\^{o}pital's rule.
The validity of Eqs.~(\ref{a-wx}) and (\ref{a-wy}) is not restricted to the vicinity of the $\Gamma$ point but holds also for large momenta. The only restriction is that the $k$-space region of interest must be away from energy anticrossings.

\begin{figure}[!h]
\centering
\includegraphics[width=0.70\columnwidth,angle=0]{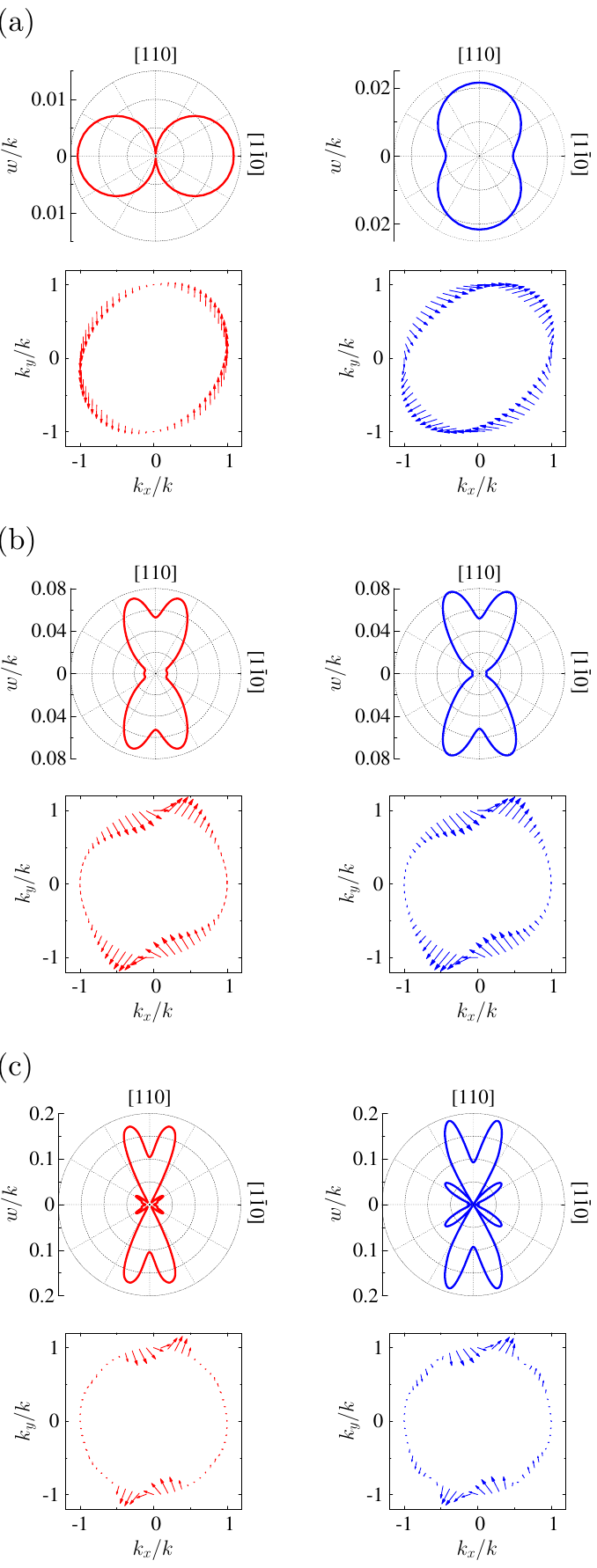}
\caption{{Spin-orbit-field ``butterflies''.} Calculated spin-orbit fields
for the magnetization along $[1\bar{1}0]$ (left) and $[110]$ (right).
The polar plots of the spin-orbit coupling strength
($w/k$) in the units of ${\rm eV\,\AA}$
as well as the corresponding vector fields $\bm{w}(\mathbf{k})$ are shown for the band $n=1$
and the momentum contours of (a)~$k=\pi/100d$;
(b)~$k=\pi/8d$;
(c)~$k=\pi/5d$.
The lengths of the direction vectors have been rescaled.
}
\label{Fig:SOF}
\end{figure}

Figure~\ref{Fig:SOF} establishes the proof of principle for the magnetization dependence of SOFs.
It shows the SOF, $\bm{w}(\mathbf{k})$ (bottom parts), and polar plots of its strength $w=|\bm{w}(\mathbf{k})|$ (upper parts), for the interface band $n=1$.
The SOF is computed on three different contours around the $\Gamma$ point, $k=\pi/100d$, $\pi/8d$ and $\pi/5d$ and plotted in Figs.~\ref{Fig:SOF}a), b), and c), respectively.
The left (right) panel corresponds to the magnetization pointing along $[1\bar{1}0]$
($[110]$). The $C_{2v}$ symmetry of the SOF is preserved for all $k$. In particular, close to the $\Gamma$ point the SOFs resemble the interference of
Bychkov-Rashba-type and Dresselhaus-type SOCs (see Fig.~\ref{Fig:SOF}a). However, away from the $\Gamma$
point higher in $k$ terms become relevant and more exotic patterns---we call them spin-orbit-field
butterflies---in the SOF appear
(see Fig.~\ref{Fig:SOF}b, c). The linear terms are dominant up to about 5\% from the BZ center,
where the SOF exhibits a very strong dependence on the magnetization orientation.
Note that the principal symmetry axes of the SOF can even be flipped by turning the magnetization orientation.
This remarkable effect opens the perspective of a magnetic control of spin-orbit fields.

Close to the $\Gamma$ point the SOF is determined by the contributions linear in the
wave vector components $k_x$ and $k_y$ and characterized by Bychkov-Rashba-type and
Dresselhaus-type SOC parameters, $\alpha_{n}=[\mu_{n}^{(0)}-\eta_{n}^{(0)}]/2$ and
$\beta_{n}=[\mu_{n}^{(0)}+\eta_{n}^{(0)}]/2$, respectively.
Using Eqs.~(\ref{a-so-expa})-(\ref{a-de-so-fp}) we obtain,
(see Supplementary Material for more details),
\begin{equation}\label{a-alpha}
    \alpha_{n}(\theta)=\sigma\left[\frac{a_{n}(\theta)\cos\theta-b_{n}(\theta)\sin\theta}{\sin(2\theta)}\right],
\end{equation}
and
\begin{equation}\label{a-beta}
       \beta_{n}(\theta)=\sigma\left[\frac{a_{n}(\theta)\cos\theta+b_{n}(\theta)\sin\theta}{\sin(2\theta)}\right],
\end{equation}
where
$a_{n}(\theta)=\left.\partial E_{n}(\mathbf{k},\theta)/\partial k_{x}\right|_{k=0}$ and
$b_{n}(\theta)=\left.\partial E_{n}(\mathbf{k},\theta)/\partial k_{y}\right|_{k=0}$.
Thus, the dependence of $\alpha_{n}(\theta)$ and $\beta_{n}(\theta)$ on the magnetization
orientation can be obtained by computing the k-space gradient (velocity) of the \emph{ab-initio}
energy bands in the vicinity of the $\Gamma$ point. The functional forms of $a_{n}(\theta)$ and
$b_{n}(\theta)$ conform to the symmetry requirements (see Supplementary Material).

Figure~\ref{Fig:SOC-param} shows the magnetization dependence of the
Bychkov-Rashba-type and Dresselhaus-type SOC parameters for the interface bands.
The SOC parameters exhibit an oscillatory behavior as a function of the magnetization orientation.
The angular dependence of the SOC parameters is stronger for band $n=1$ than for $n=2$. In particular,
for the case of band $n=1$ the Bychkov-Rashba-type SOC parameter can even change its sign when the magnetization is rotated in the plane.
This leads to the sign change of the product $\alpha_{1}\beta_{1}$ when the magnetization is rotated from $[1\bar{1}0]$ to $[110]$
and produces the flipping of the
SOF symmetry axes shown in Fig.~\ref{Fig:SOF}a).
For  band $n=2$ the angular dependence is weaker, the product $\alpha_{2}\beta_{2}$ does not change its sign
and the symmetry axis of the SOF is preserved, being independent of the magnetization orientation.

When considering the
dependence on the transverse electric field, the behavior is opposite. Indeed, while the SOC parameters corresponding to band $n=1$
change very little with $E$, for band $n=2$ the changes in the magnitudes of $\alpha_2$ and $\beta_2$ are appreciable.
This disparate behavior is a consequence of the different nature of these two bands. Band $n=1$ originates mostly from the two Fe
atoms in the interface unit cell and, therefore, its corresponding SOF is more sensible to the changes in the magnetization direction.
However, the electrostatic control of the SOC parameters is dominated by the
electric field influence on the $pd$ bonding between As and Fe atoms.
Consequently, the SOF corresponding to  band $n=2$, which comes mostly from the
interfacial As atom, its neighboring Ga and the Fe atom above Ga, exhibits
a stronger dependence on the electric field.
\begin{figure}[!h]
\centering
\includegraphics[width=0.99\columnwidth,angle=0]{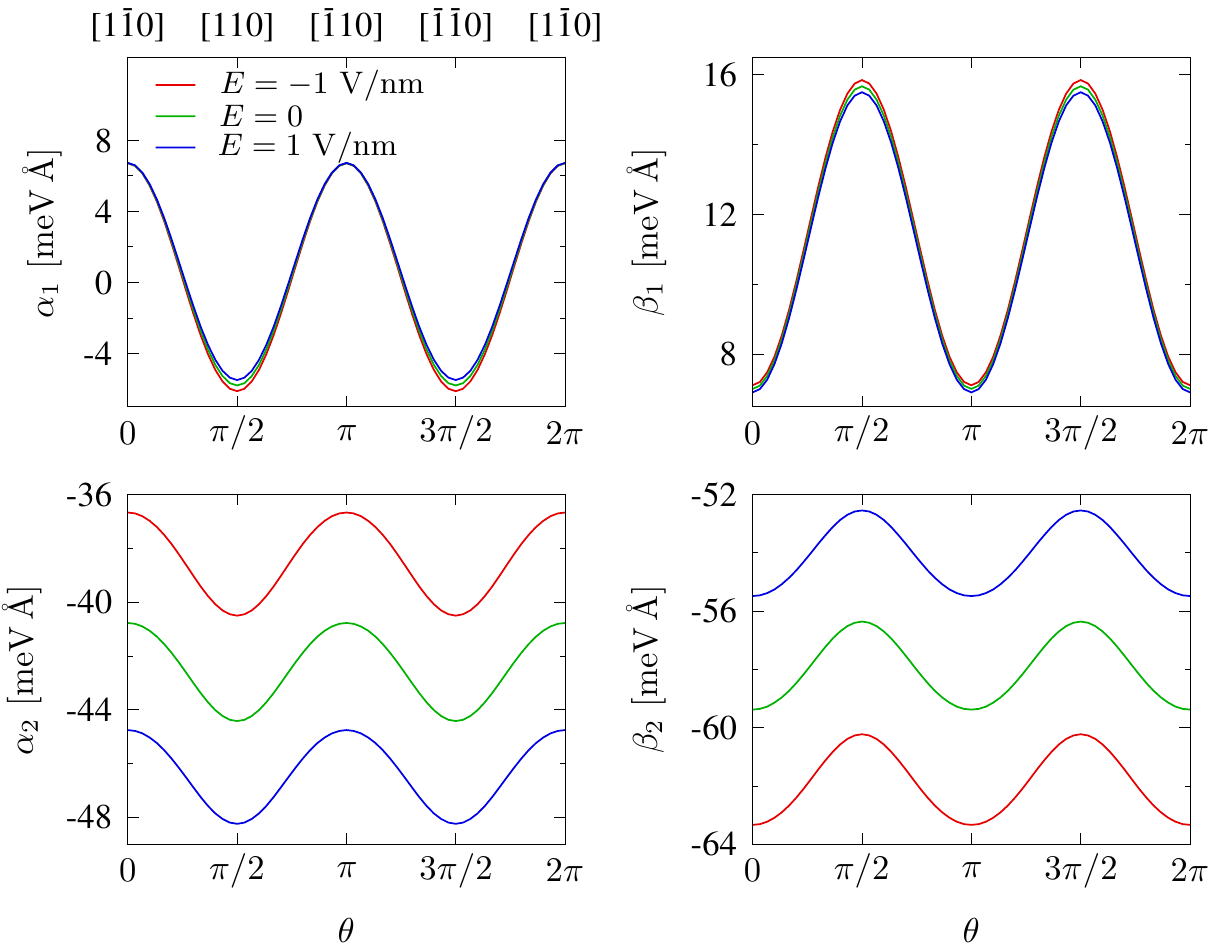}
\caption{{Calculated magnetization and electric field dependence of the spin-orbit
coupling parameters.} The Bychkov-Rashba-type $\alpha_{n}$ and the Dresselhaus-type
$\beta_{n}$ spin-orbit parameters for the interface bands $n=1,\,2$ are shown as a function
of the in-plane magnetization orientation and for different electric fields.
}
\label{Fig:SOC-param}
\end{figure}

In Table~I we list the expansion coefficients of the SOC parameters [see Eqs.~(23-24) in the Supplementary Material],
\begin{eqnarray}
\alpha_n & \simeq & A_n^{(+)} + B_n^{(+)}\cos(2\theta), \\
\beta_n &\simeq & A_n^{(-)} + B_n^{(-)}\cos(2\theta),
\end{eqnarray}
 for zero electric field. From $A_n^{(+/-)}$ one extracts the magnetization-independent part, whereas
the $B_n^{(+/-)}$ parameters control the leading contribution (higher order coefficients are about two orders smaller) to
the angular (magnetization orientation) dependence of the spin-orbit parameters. In addition to the interface bands ($n=1,\;2$) we have also included the expansion coefficients corresponding to the As-surface bands ($n=3,\;4$), which due to their surface nature possess stronger SOFs.
%
\begin{table}[!h]\centering
\begin{minipage}{\columnwidth}
\begin{tabular}{@{\extracolsep{1.5em}}crrrr}
$n$ & $A_n^{(+)}$ & $B_n^{(+)}$ & $A_n^{(-)}$ & $B_n^{(-)}$ \\
\hline\hline
1 &  -0.42 & -6.26 & -11.32 & 4.32 \\
2 & -42.51 & 1.82  & -57.94 & -1.51 \\
3 & -620.24 & -88.74 & -597.56 & -89.62 \\
4 & 680.09 & 95.61 & 697.58 & 103.15 \\
\hline
\end{tabular}
\caption{{Band-resolved expansion coefficients of the Bychkov-Rashba-type and Dresselhaus-type
spin-orbit coupling parameters in ${\rm meV\,\AA}$ units. The parameters are in the range
of what is found in  semiconductors \cite{Fabian2007:APS}), }
} \label{Tab:BRD}
\end{minipage}
\end{table}

\begin{figure}[!h]
\centering
\includegraphics[width =0.85\columnwidth,angle=0]{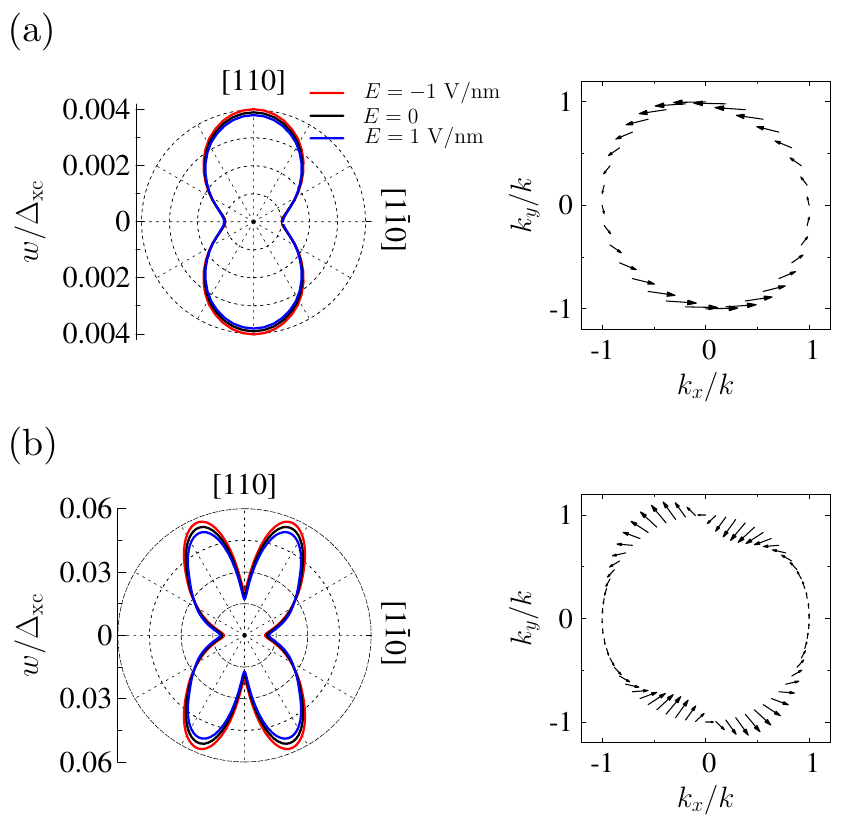}
\caption{{Electric field control of the spin-orbit coupling for magnetization
perpendicular to the plane.} Polar plots of the spin-orbit field strength (left panel) corresponding to the interface band $n=1$ on the contours (a)~$k=\pi/25d$ and (b)~$k=\pi/8d$, for transverse electric
fields -1, 0, 1~V/nm. The momentum-dependent expectation values of the in-plane spin vectors $\langle\bm{s}\rangle$ on the two corresponding $k$-contours
is shown in the right panel for zero electric field. The size of the vectors has been rescaled.}
\label{Fig:SOF_m001}
\end{figure}

If the magnetization is perpendicular to the interface plane,  the first order
correction to the energy vanishes and the methodology used for the case
of in-plane magnetization does not apply. To extract useful information
about  SOFs we first note that the in-plane components of the spin
appear due to SOFs only (without spin-orbit coupling the electron spins
would be fully polarized in the growth direction).
Since the exchange field dominates over  SOC, the spin is still quantized
largely along the magnetization direction, so the expectation values
of the transverse components of the spin $\langle\bm{s}\rangle_{n}$
corresponding to the $n$th band can be obtained by considering
${\cal H}_{\rm so}$ as a perturbation. First order perturbation theory gives
\begin{equation}
\langle s_x \rangle_{n} = w_{nx}/\Delta_{\rm xc}
\qquad ; \qquad
\langle s_y \rangle_{n} = w_{ny}/\Delta_{\rm xc} \,,
\end{equation}
where $\Delta_{\rm xc}$ is the exchange splitting energy and $\hbar=1$.
Using these approcimate relations, we can determine the pattern of $\bm{w}$, but
not its magnitude.
Figure~\ref{Fig:SOF_m001} shows  $\bm{w}(\mathbf{k})$ (right parts), and its rescaled magnitude
$w=|\bm{w}(\mathbf{k})|$ (left parts) in the units of the exchange splitting $\Delta_{\rm xc}$ for the interface band $n=1$.
The fields have been computed on two contours, $k=\pi/25d$ (a) and $k=\pi/8d$ (b).
Similar to the in-plane magnetization case, when the magnetization is perpendicular to the layers, the SOF close to the $\Gamma$
point  resembles the interference of the Bychkov-Rashba-type and
Dresselhaus-type SOCs. The effect of a transverse electric field quantitatively
modifies the SOF and is more pronounced for larger $k$ values [see Fig.~\ref{Fig:SOF_m001}b)].
%

To summarize, we introduced a method to calculate spin-orbit fields 
from first principles and applied it to the Fe/GaAs interface. 
We found the the spin-orbit fields depend strongly on the
Fe magnetization direction. This finding should be 
particularly important for lateral and tunneling
magnetotransport anisotropies of ferromagnet-nonmagnet slabs.

We thank D.~Weiss, G.~Bayreuther, C.~Back, G.~Woltersdorf
for useful hints related to experimental ramifications of the
presented theoretical concepts and F.~Freimuth, Y.~Mokrousov,
G.~Bihlmayer, J.~Spitaler and P.~Nov\'ak for helpful discussions
regarding the calculations. This work has been supported by DFG SFB 689.

\bibliography{spin}

\clearpage

\pagestyle{empty}

\begin{widetext}

\includegraphics[width=1.0\columnwidth,angle=0]{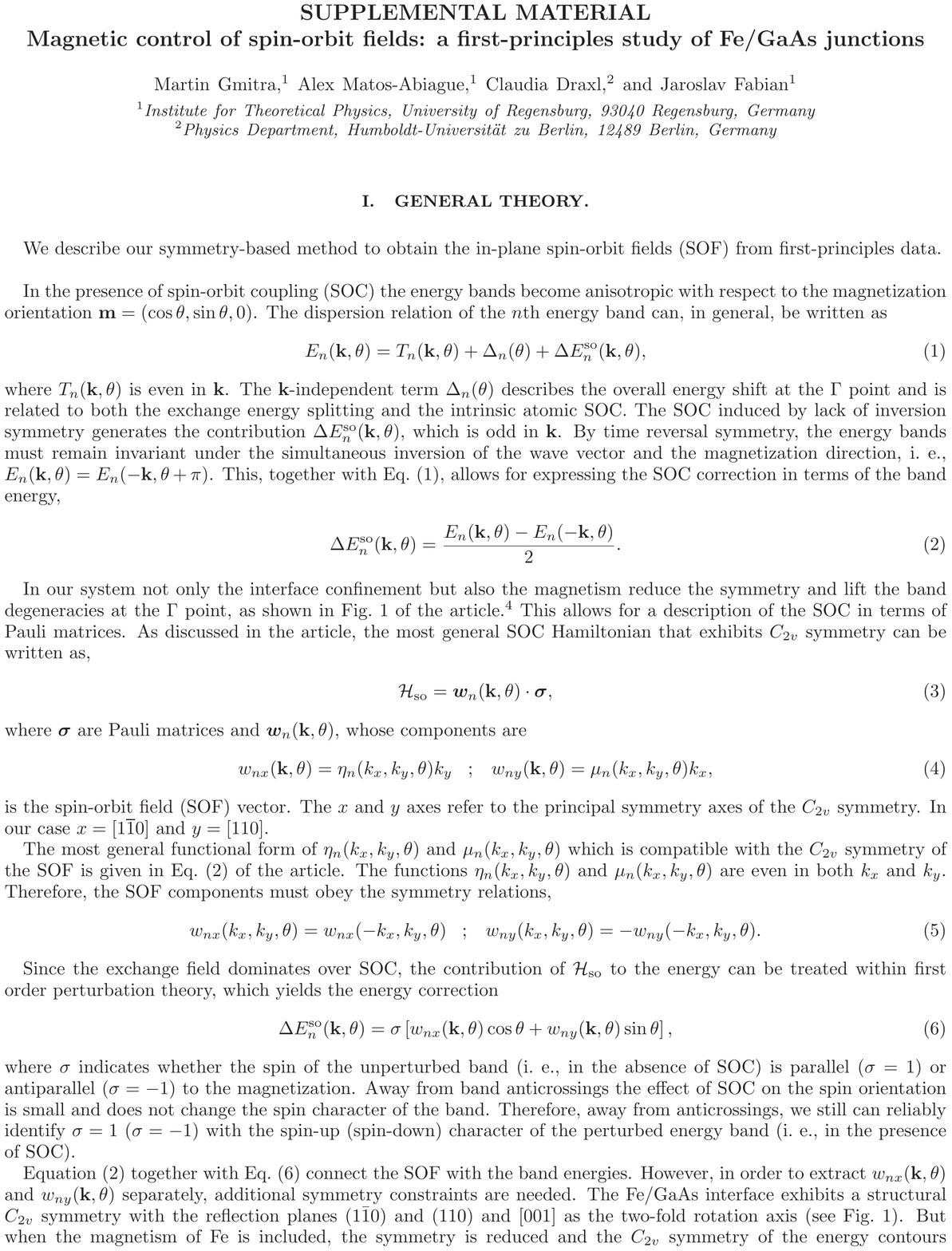}
\includegraphics[width=1.0\columnwidth,angle=0]{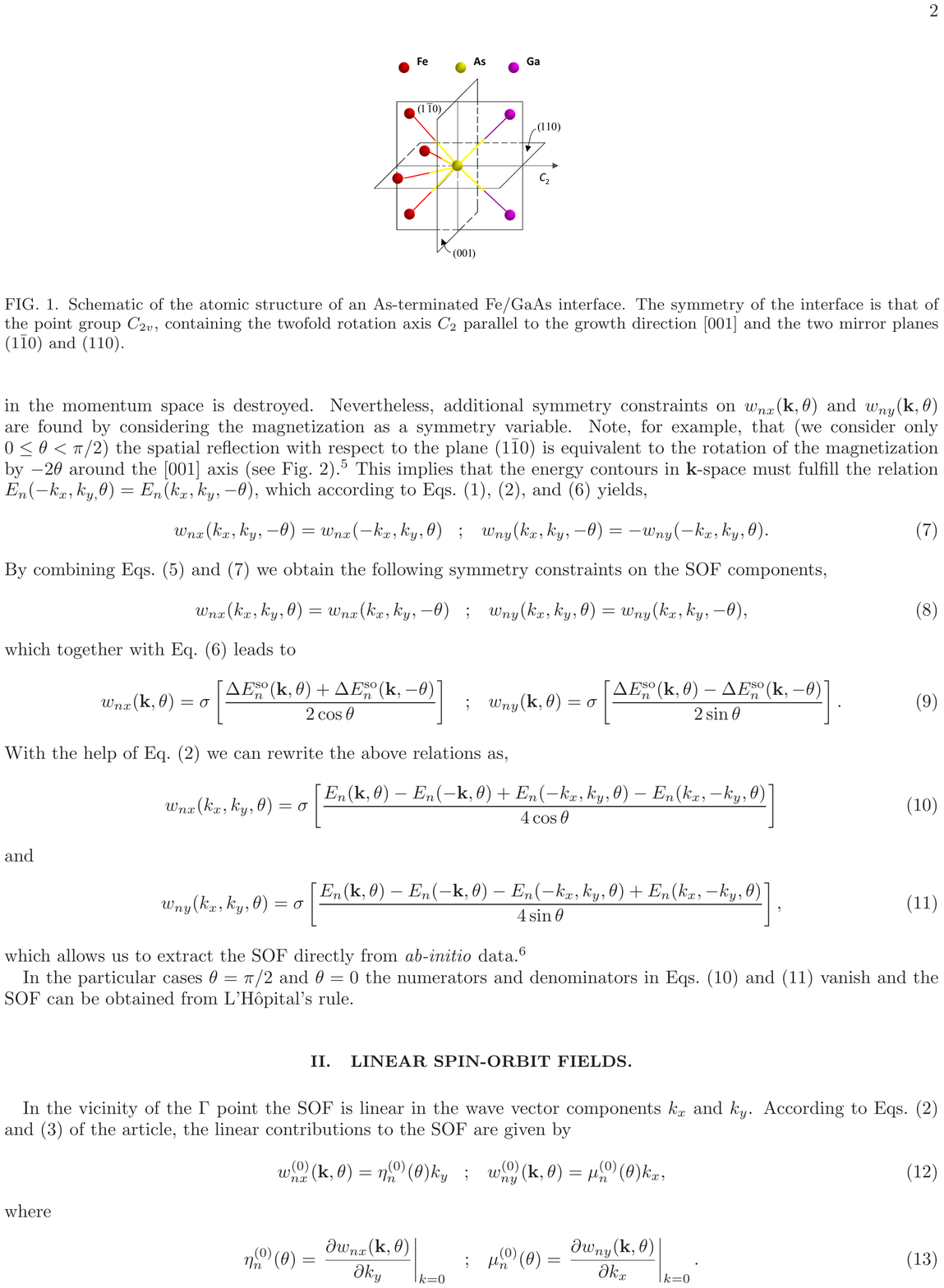}
\includegraphics[width=1.0\columnwidth,angle=0]{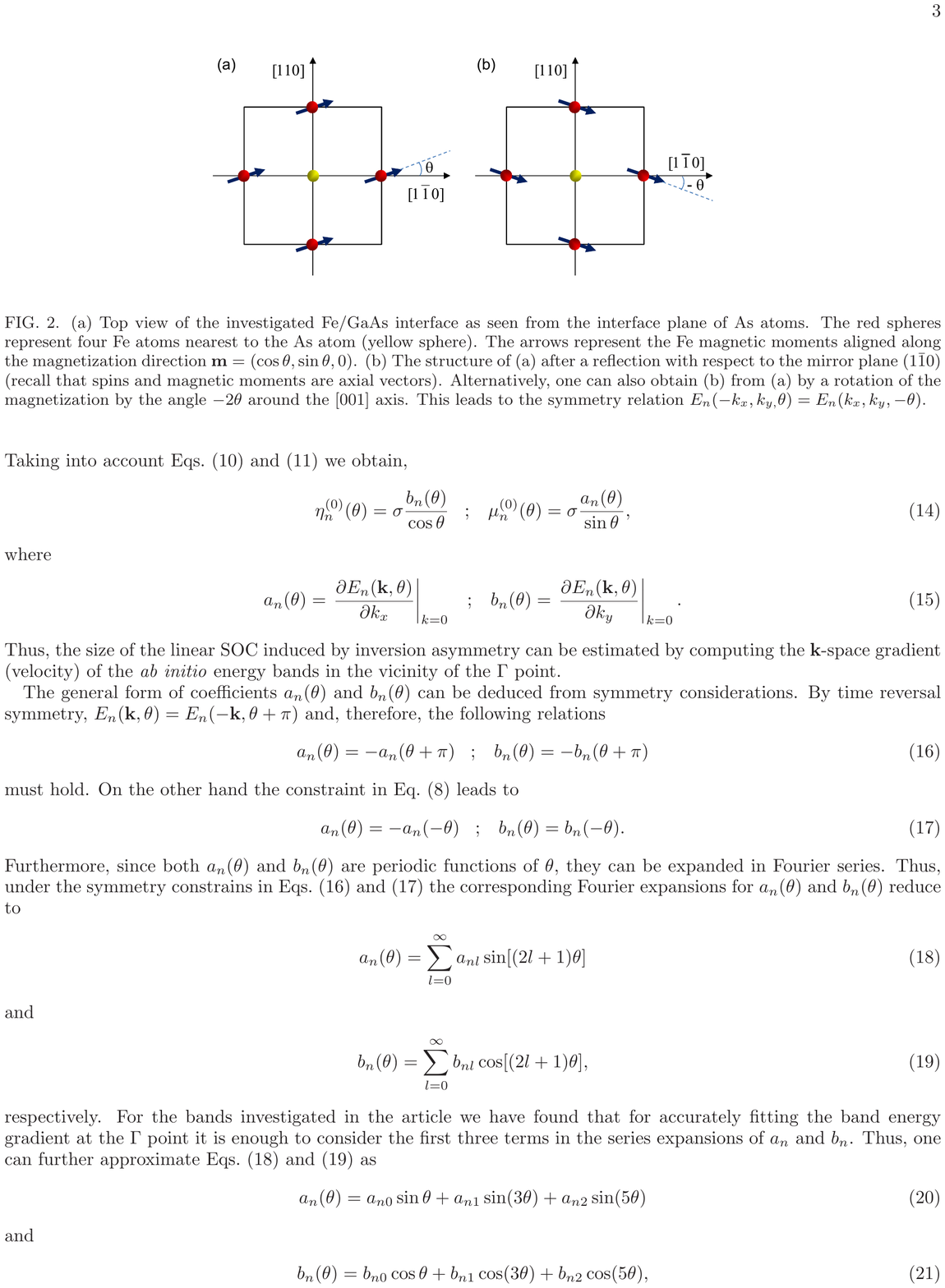}
\includegraphics[width=1.0\columnwidth,angle=0]{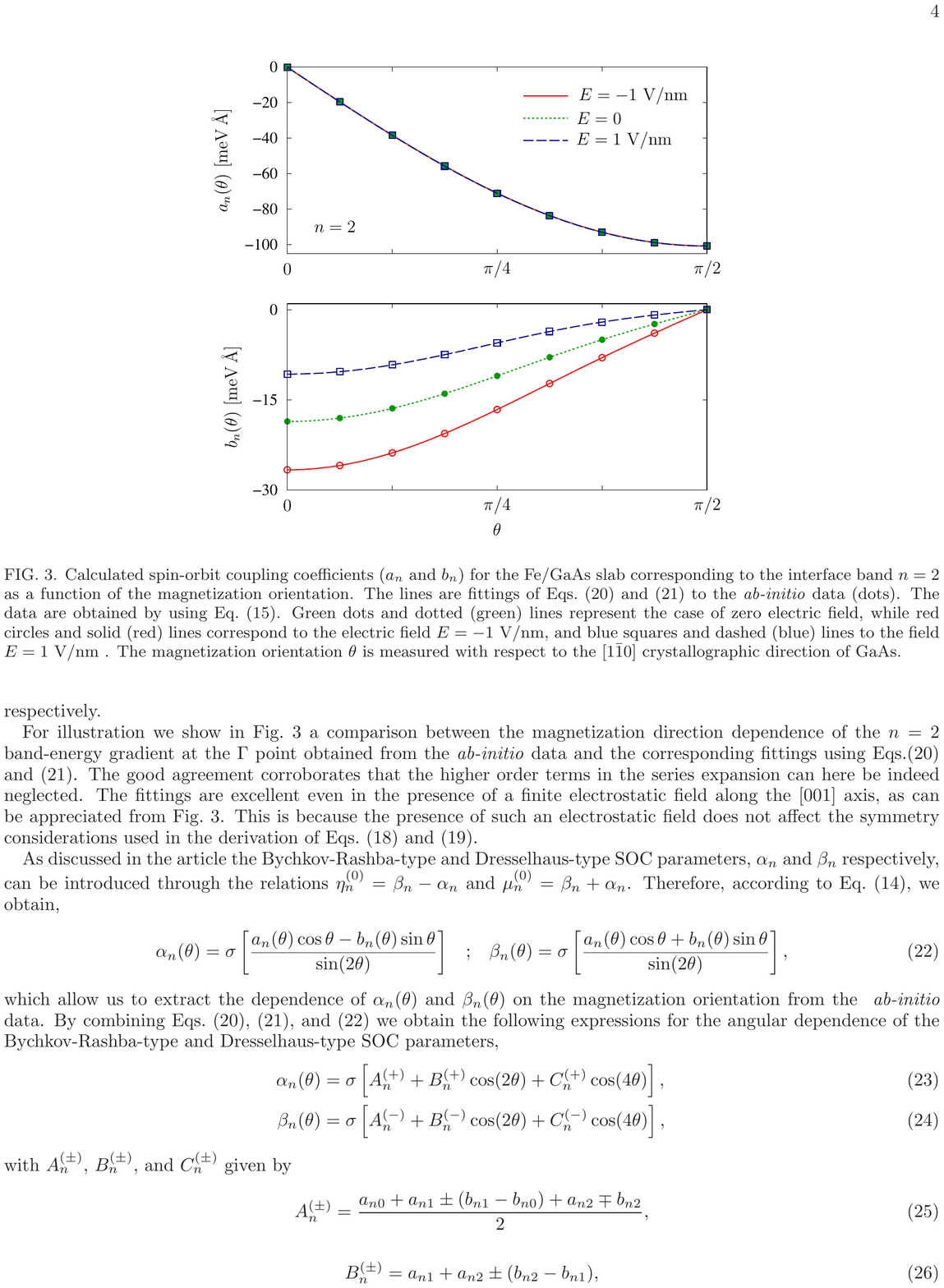}
\includegraphics[width=1.0\columnwidth,angle=0]{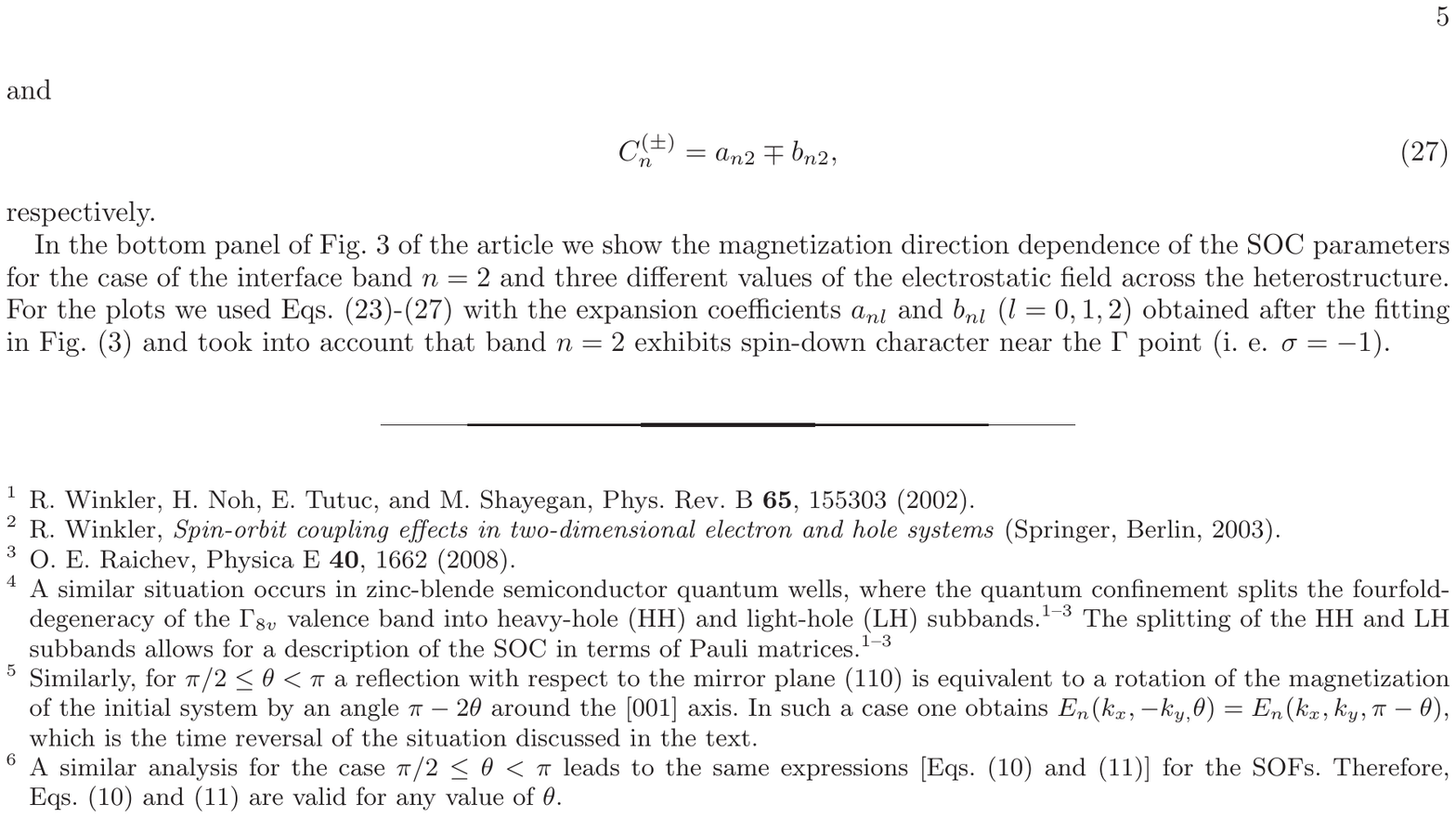}

\end{widetext}

\end{document}